# Extrinsic nonlinear Kerr rotation in topological materials under a magnetic field

Shuang Wu[1], Zaiyao Fei[2], Zeyuan Sun[1], Yangfan Yi[1], Wei Xia[3], Dayu Yan[4], Yanfeng Guo[3], Youguo Shi[4], Jiaqiang Yan[5], David H. Cobden[2], Wei-Tao Liu[1], Xiaodong Xu[2,6], Shiwei Wu[1,7,8,9*]

[1] *State Key Laboratory of Surface Physics, Key Laboratory of Micro and Nano Photonic Structures (MOE), and Department of Physics, Fudan University, Shanghai 200433, China.*
[2] *Department of Physics, University of Washington, Seattle, Washington 98195, USA*
[3] *School of Physical Science and Technology, and ShanghaiTech Laboratory for Topological Physics, ShanghaiTech University, Shanghai 201210, China*
[4] *Beijing National Laboratory for Condensed Matter Physics, Institute of Physics, Chinese Academy of Sciences, Beijing 100190, China*
[5] *Materials Science and Technology Division, Oak Ridge National Laboratory, Oak Ridge, Tennessee, 37831, USA*
[6] *Department of Materials Science and Engineering, University of Washington, Seattle, Washington 98195, USA*
[7] *Shanghai Qi Zhi Institute, Shanghai 200232, China*
[8] *Institute for Nanoelectronic Devices and Quantum Computing, and Zhangjiang Fudan International Innovation Center, Fudan University, Shanghai 200433, China*
[9] *Shanghai Research Center for Quantum Sciences, Shanghai 201315, China*

* Corresponding emails: swwu@fudan.edu.cn

**Abstract**

Topological properties in quantum materials are often governed by symmetry and tuned by crystal structure and external fields, and hence symmetry-sensitive nonlinear optical measurements in a magnetic field are a valuable probe. Here we report nonlinear magneto-optical second harmonic generation (SHG) studies of non-magnetic topological materials including bilayer $WTe_2$, monolayer $WSe_2$ and bulk TaAs. The polarization-resolved patterns of optical SHG under magnetic field show nonlinear Kerr rotation in these time-reversal symmetric materials. For materials with three-fold rotational symmetric lattice structure, the SHG polarization pattern rotates just slightly in a magnetic field, whereas in those with mirror or two-fold rotational symmetry the SHG polarization pattern rotates greatly and distorts. These different magneto-SHG characters can be understood by considering the superposition of the magnetic field-induced time-noninvariant nonlinear optical tensor and the crystal-structure-based time-invariant counterpart. The situation is further clarified by scrutinizing the Faraday rotation, whose subtle interplay with crystal symmetry accounts for the diverse behavior of the extrinsic nonlinear Kerr rotation in different materials. Our work illustrates the application of magneto-SHG techniques to directly probe nontrivial topological properties, and underlines the importance of minimizing extrinsic nonlinear Kerr rotation in polarization-resolved magneto-optical studies.

**Main text**

The discovery of Berry curvature and energy band topology provides an exclusive route to generate topologically protected states.[1-6] Geometric effect of the energy band is proposed to be essential for understanding a series of fascinating phenomena, such as valley magnetization in semiconducting transition metal dichalcogenides,[7-11] edge conduction in magnetic and 2D topological insulators,[12-14] negative magnetoresistance in topological semimetals[15, 16] and nonlinear Hall effect in Weyl semimetals.[17-19] In these interesting phenomena, symmetry often plays the key role in determining the band structure and topology.

Nonlinear optical processes such as second harmonic generation (SHG) are sensitive to symmetry[20, 21] and have been proposed to study the topological effects.[22] Great efforts indeed have been made to search for the topological nonlinear optical effects. For example, extraordinary strong SHG was observed in topological Weyl semimetal TaAs and interpreted as the consequence of Berry curvature enhancement.[23] Circular photogalvanic effect was also used to probe the topological charge of Weyl points.[24-27] Linear photovoltaic effect was observed in TaAs and associated with the Berry connection and curvature of the Bloch bands.[26, 28] While these nonlinear optical effects are intriguing, their connection with band topology is still indirect. To examine their topological nature, it would be desirable to employ some kind of symmetry-tuning knobs to alter the band topology and then in-situ probe the corresponding change of nonlinear optical effects.

In this paper, we apply the external magnetic field to non-magnetic topological materials and investigate the optical second harmonic generation effect associated with field-induced time-reversal symmetry breaking. In time-reversal symmetric topological materials, the magnetic field can induce strong orbital magnetization and generate the time-noninvariant second-order nonlinear optical response. The superposition of emergent time-noninvariant nonlinear optical tensor with the crystal-structure-based time-invariant counterpart would modify the polarization-resolved SHG, which exhibits as the nonlinear Kerr rotation in polarization patterns. Here we experimentally search for the nonlinear Kerr rotation in several topological materials, including $WTe_2$ bilayer, $WSe_2$ monolayer and 3D TaAs crystal. By studying the magneto-SHG in Faraday geometry, giant rotation and distortion of SHG polarization patterns are observed in $WTe_2$ bilayer and TaAs, while only a relatively small rotation is observed in $WSe_2$ monolayer. The contrasting behaviors from different materials with distinct crystal symmetry are understood by scrutinizing the Faraday rotation effect through microscope objective housed in magnetic field. The observed magneto-SHG is thus dominated by Faraday rotation effect and the intrinsic nonlinear Kerr rotation remains absent. We highlight the importance of minimizing the extrinsic effect, and suggest several routes to enhance the intrinsic nonlinear Kerr rotation.

**SHG method**

We used a home-built magneto-optical cryostat to conduct SHG measurement, as shown in Figure 1a. The sample was mounted into the cryostat housed inside a superconducting magnet with a room-temperature bore. Femtosecond laser pulses from a Ti:Sapphire oscillator (MaiTai HP, Spectra Physics) at optical frequency ω were focused onto the sample by a microscope objective at normal incidence. The reflected beam was collected by the same objective. The fundamental light was blocked by a band-pass filter and the frequency-doubled SHG signal was detected by a photomultiplier tube in photon counting mode. The XX or XY polarization-resolved SHG patterns were obtained by setting the fundamental and signal beams to co- or cross-linearly polarized, respectively, with a half wave plate and a linear polarizer. Because of the small sample size (just a few microns in length), the data points of polarization-resolved SHG patterns and magnetic field dependent SHG signals were obtained from series of polarization-resolved SHG images at different azimuthal angles. To save time, we measured the polarization-resolved SHG only up to 180° azimuthal angle, and projected the data at $\theta$ to that at $\theta + 180°$. This protocol was validated for a complete 360° rotation of the azimuthal angle. We extracted the nonlinear Kerr rotation angle from the obtained magneto-SHG patterns.

**Nonlinear Kerr rotation**

We first analyze the topological properties of non-magnetic crystals in momentum space. For a time-reversal symmetric system, the inversion symmetry must be broken to allow nonvanishing orbital magnetic moment and Berry curvature. Taking the 2D materials as an example (Figure 1a), broken spatial-inversion symmetry gives rise to momentum dependent orbital magnetic moment $\mathbf{m_z}$ and Berry curvature $\mathbf{\Omega_z}$. Under the constraint of the time-reversal symmetry, $\mathbf{m_z}$ takes opposite sign at opposite momentum **k,** which guarantees the vanishing of total magnetic moment summing over the entire Brillouin zone. When an external magnetic field is applied along the out-of-plane direction, the time-reversal symmetry is broken and thus the degeneracy of electronic bands at opposite **k** would be lifted due to the $\mathbf{m} \cdot \mathbf{B}$ term, as illustrated in Figure 1b. Furthermore, the Berry phase correction[29] due to $(1 + e\mathbf{B} \cdot \mathbf{\Omega}/\hbar)^{-1}$ could modify the electron density of states.

Nonlinear optical responses such as SHG are known to be a powerful probe for various symmetries. For example, SHG is sensitive to the inversion symmetry breaking under electric dipole approximation, and has been widely applied to study the symmetry and symmetry breaking in quantum materials.[30, 31] According to the Neumann's principle, symmetry determines the nonlinear optical tensor elements, which could be measured through polarization-resolved SHG. For a material system with time-reversal symmetry, the second-order nonlinear optical tensor is solely determined by the time-invariant part $\chi^{(i)}$ (i-type). When the magnetic field is applied to break the time-reversal symmetry and induce a net magnetization, the topological nonlinear optical effect could be manifested through the emergent time-noninvariant part $\chi^{(c)}$ (c-type). The coherent superposition of $\chi^{(i)}$ and $\chi^{(c)}$ would contribute to the total SHG signal. Such tensor-based superposition is much like vector addition. It will lead to the tilt of second-order polarization direction, experimentally manifested as the rotation of polarization-resolved SHG patterns. The magnetization induced rotation of SHG patterns is called nonlinear Kerr rotation, and has been observed in ferromagnetic materials.[32-34] Because $\chi^{(c)}$ is odd and $\chi^{(i)}$ is even to the magnetic field, the flip of magnetic field direction would lead to the opposite rotation of SHG patterns. For topological materials discussed here, the nonlinear Kerr rotation is a result of the interaction between band topology and applied magnetic field.

Apart from the intrinsic nonlinear Kerr rotation, where topological nonlinear optical effect occurs on the second-order electric polarization, some extrinsic magneto-optical effects may also play the role in such magneto-SHG measurements. In particular, the optical components under magnetic field often induce Faraday rotation when the magnetic field is parallel to the propagation direction of light. Such the first-order modification to light polarization will affect both the fundamental beam and second harmonic signal, and cause an extrinsic nonlinear Kerr rotation. Thus, the scrutiny of nonlinear Kerr rotation is essential to searching for topological nonlinear optical effects.

**WTe$_2$ bilayer with mirror symmetry**

To investigate such nonlinear Kerr rotation, we firstly study a two-dimensional topological semimetal WTe$_2$.[18, 19, 35] Because odd-layer WTe$_2$ is centrosymmetric and even-layer is non-centrosymmetric, we choose WTe$_2$ bilayer to study the nonlinear Kerr rotation. Figure 2a shows the atomic structure of WTe$_2$ bilayer in 1T' phase that is non-centrosymmetric with a mirror plane (marked by the black dashed line) normal to the a-axis. Figure 2b shows a white-light optical microscopy of a WTe$_2$ bilayer sample prepared by mechanical exfoliation on SiO$_2$/Si substrate. Its thickness is identified based on their optical contrasts. Since SHG is sensitive to inversion symmetry breaking, the WTe$_2$ bilayer region emits strong SHG signal as shown in Figure 2c, excited by a femtosecond laser at wavelength of 800 nm.

The strong SHG in WTe$_2$ bilayer permits the extraction of structural information such as the mirror plane through the polarization-resolved measurement. Figure 2d shows the polarization-resolved SHG. The dumbbell-shape XX and XY SHG patterns are fitted in solid lines by considering the electric-dipole contribution and the $C_{1h}$ symmetry of WTe$_2$ bilayer. There are three independent nonzero i-type ($\chi^{(i)}$) tensor elements:[20] $\chi^{(i)}_{bbb}, \chi^{(i)}_{baa}, \chi^{(i)}_{aab} = \chi^{(i)}_{aba}$. The relation between the input and output electric fields is:

$$E_{2\omega}^{\parallel} \propto \begin{pmatrix} \chi^{(i)}_{bbb}\cos^3(\varphi+\theta_0) \\ +(\chi^{(i)}_{baa} + 2\chi^{(i)}_{aab})\sin^2(\varphi+\theta_0)\cos(\varphi+\theta_0) \end{pmatrix} E_\omega^2 \qquad (1)$$

$$E_{2\omega}^{\perp} \propto \begin{pmatrix} \chi^{(i)}_{baa}\sin^3(\varphi+\theta_0) \\ +(\chi^{(i)}_{bbb} - 2\chi^{(i)}_{aab})\sin(\varphi+\theta_0)\cos^2(\varphi+\theta_0) \end{pmatrix} E_\omega^2 \qquad (2)$$

$E_{2\omega}^{\parallel}$ and $E_{2\omega}^{\perp}$ are the second harmonic output electric fields in the XX and XY polarization configurations, respectively. $E_\omega$ is the electric field of incident beam. $\varphi$ indicates the azimuthal angle of incident beam, and $\theta_0$ indicates the angle between the crystal axis and the initial azimuth of incident beam. This fitting is used to determine the crystal axes of the sample.

To study the nonlinear magneto-optical response of WTe$_2$ bilayer, we applied an external out-of-plane magnetic field, while maintaining the zero-field optical settings. Figure 2e shows the magneto-SHG patterns of the WTe$_2$ bilayer in out-of-plane magnetic field from -4 T to 4 T. For a direct comparison, the zero-field patterns in Figure. 2d are also plotted by the dashed lines and filled in light colors. At negative magnetic field, the magneto-SHG patterns rotate counter-clockwise and also distort in shape. Reversing the direction of magnetic field leads to the clockwise rotation of patterns and the similar distortion. The rotation in magneto-SHG patterns is similar to those phenomena previously found in ferromagnetic materials,[32-34] suggesting the existence of nonlinear Kerr rotation in WTe$_2$ bilayer.

### WSe$_2$ monolayer with $C_3$ symmetry

For comparison, we carried out the same magneto-SHG measurement on mechanical exfoliated WSe$_2$ monolayer, which belongs to a higher symmetry point group ($D_{3h}$). As the out-of-plane magnetic field could interact with the non-zero orbital magnetic moment and induce valley/spin polarization, nonlinear Kerr rotation may also appear. As shown in Figure 3a, under ±4 T magnetic field, the SHG patterns rotate counterclockwise and clockwise, respectively. Beside rotating, the shape of patterns remains unchanged, and the rotation angle is surprisingly several times smaller than that in WTe$_2$.

The zero-field polarization-resolved SHG patterns were fitted by considering the $D_{3h}$ point group of its crystalline structure, which has one independent nonzero tensor element: $\chi^{(i)}_{bbb} = -\chi^{(i)}_{baa} = -\chi^{(i)}_{aab} = -\chi^{(i)}_{aba}$. The fitting formula is:

$$E_{2\omega}^{\parallel} \propto \chi^{(i)}_{bbb} \cos 3(\varphi + \theta_0) E_{\omega}^2 \tag{3}$$

$$E_{2\omega}^{\perp} \propto \chi^{(i)}_{bbb} \sin 3(\varphi + \theta_0) E_{\omega}^2 \tag{4}$$

$\theta_0$ indicates the angle between the crystal axis and the initial azimuth of incident beam. Since no magnetic field induced distortion is found, the same formula is used to fit the patterns under external magnetic field. The angle $\theta_0$ shows a linear relation with the magnetic field (Figure 3b), and the corresponding slope is 2.2 °/T.

### Data fitting and interpretation

As seen, the behavior of nonlinear Kerr rotation induced by external magnetic field is different between WT$_2$ bilayer and WSe$_2$ monolayer. This difference might be related with their distinct crystallographic symmetry point group ($C_{1h}$ for WTe$_2$ bilayer versus $D_{3h}$ for WSe$_2$ monolayer). We then analyze the nonlinear optical tensor elements based on the symmetry. For time invariant i-type tensor $\chi^{(i)}$, the non-zero tensor elements are determined by their crystallographic point group. When the magnetic field is applied, the time-noninvariant c-type tensor $\chi^{(c)}$ becomes non-zero and linearly depends on magnetic field. The non-zero tensor elements of $\chi^{(c)}$ are deduced from the corresponding magnetic point group. As shown below, the non-zero tensor elements of $\chi^{(i)}$ and $\chi^{(c)}$ are paired in orthogonal direction. The superposition of $\chi^{(i)}$ and $\chi^{(c)}$ would lead to the tilt of second-order polarization direction and the rotation of polarization-resolved SHG patterns (Figure 4). Because the flip of magnetic field direction changes the sign of $\chi^{(c)}$ and keeps the sign of $\chi^{(i)}$, the SHG polarization patterns at opposite magnetic fields rotate reversely and form the enantiomorphous counterparts.

For WTe$_2$ bilayer, the crystallographic symmetry point group is $C_{1h}$ (*m*) that contains mirror and identity symmetry operation. The non-zero elements of i-type $\chi^{(i)}$ tensor are $\chi^{(i)}_{bbb}$, $\chi^{(i)}_{baa}$ and $\chi^{(i)}_{aab}$. The corresponding magnetic point group is ***m*** and thus the non-zero elements of c-type $\chi^{(c)}$ tensor are $\chi^{(c)}_{abb}$, $\chi^{(c)}_{aaa}$ and $\chi^{(c)}_{bab}$. By projecting $\chi^{(i)}$ and $\chi^{(c)}$ from the crystallographic coordinates to the laboratory coordinates, the fitting formula for the magneto-SHG patterns is deduced:

$$E_{2\omega}^{\parallel} \propto \begin{bmatrix} \chi^{(i)}_{bbb}\cos^3(\varphi+\theta_0) \\ +(\chi^{(i)}_{baa}+2\chi^{(i)}_{aab})\sin^2(\varphi+\theta_0)\cos(\varphi+\theta_0) \\ +\chi^{(c)}_{aaa}\sin^3(\varphi+\theta_0) \\ +(\chi^{(c)}_{abb}+2\chi^{(c)}_{bab})\sin(\varphi+\theta_0)\cos^2(\varphi+\theta_0) \end{bmatrix} E_\omega^2 \qquad (5)$$

$$E_{2\omega}^{\perp} \propto \begin{bmatrix} \chi^{(i)}_{baa}\sin^3(\varphi+\theta_0) \\ +(\chi^{(i)}_{bbb}-2\chi^{(i)}_{aab})\sin(\varphi+\theta_0)\cos^2(\varphi+\theta_0) \\ +\chi^{(c)}_{abb}\cos^3(\varphi+\theta_0) \\ +(2\chi^{(c)}_{bab}-\chi^{(c)}_{aaa})\sin^2(\varphi+\theta_0)\cos(\varphi+\theta_0) \end{bmatrix} E_\omega^2 \qquad (6)$$

The fitting results of SHG patterns under $\pm 4$ T are shown in Figure 4b. The rotation and distortion of the SHG patterns are well reproduced, as the characteristics of nonlinear Kerr rotation for WTe$_2$ bilayer with mirror symmetry. The deduced nonlinear optical tensor elements as a function of magnetic field are plotted in Figure 4c, exemplified by $\chi^{(i)}_{baa}$ and $\chi^{(c)}_{aaa}$. The c-type tensor element $\chi^{(c)}_{aaa}$ shows the linear dependence of the magnetic field. In contrast, the corresponding i-type tensor element $\chi^{(i)}_{baa}$ that is orthogonal to $\chi^{(c)}_{aaa}$ remains constant.

For WSe$_2$ monolayer with $D_{3h}$ symmetry point group, similar analysis could be applied. The deduced fitting formula is:

$$E_{2\omega}^{\parallel} \propto [\chi^{(i)}_{bbb}\cos 3(\varphi+\theta_0) + \chi^{(c)}_{abb}\sin 3(\varphi+\theta_0)]E_\omega^2 \qquad (7)$$

$$E_{2\omega}^{\perp} \propto [\chi^{(c)}_{abb}\cos 3(\varphi+\theta_0) + \chi^{(i)}_{bbb}\sin 3(\varphi+\theta_0)]E_\omega^2 \qquad (8)$$

where $\chi^{(i)}_{bbb}$ is the only non-zero independent i-type tensor element, and $\chi^{(c)}_{abb}$ is the orthogonal c-type tensor element. The comparison to the zero-field case (Equation 3&4) shows that the appearance of c-type nonlinear optical tensor simply causes the rotation of six-fold anisotropic SHG pattern. We then use the above formula to fit the SHG polarization patterns in Figure 3a and obtain the magnetic field dependent $\chi^{(i)}_{bbb}$ and $\chi^{(c)}_{abb}$ (Figure 4d). Again, the c-type tensor element $\chi^{(c)}_{abb}$ is linearly dependent on magnetic field, while the i-type tensor element $\chi^{(i)}_{bbb}$ is constant.

The fitting for WTe$_2$ bilayer and WSe$_2$ monolayer demonstrates that nonlinear Kerr rotation is the cause of the magneto-SHG patterns, in which the magnetic field dependent $\chi^{(c)}$ superposes with the orthogonal $\chi^{(i)}$ and induces the rotation (and distortion) of SHG patterns. The non-zero elements of $\chi^{(i)}$ and $\chi^{(c)}$ are determined by the symmetry point group of the crystal, which is distinct for WTe$_2$ bilayer ($C_{1h}$) and WSe$_2$ monolayer ($D_{3h}$). Thus, for materials with various symmetry point group, their manifestation of nonlinear Kerr rotation behaves differently.

Besides the intrinsic topological nonlinear optical effect, nonlinear Kerr rotation could also be induced extrinsically, such as Faraday rotation effect in the beam path. For example, fused silica, one of the common optical materials for a microscope objective, has a considerable Verdet constant (about 1.5 °/T·cm at wavelength of 800 nm).[36] When the fundamental and second harmonic light beams pass through the microscope objective under magnetic field, the Faraday rotation angle for a centimeter-thick glass is appreciable, and causes extrinsic nonlinear Kerr rotation.

We next scrutinize the effect of Faraday rotation through the microscope objective and resolve the origin of nonlinear Kerr rotation. In our experimental setup (Figure 1a), the microscope objective is housed inside the room-temperature bore of superconducting magnet, and the propagation direction of light is parallel to the magnetic field. Because of the Faraday rotation effect, both the fundamental and second harmonic beams rotate their linear polarizations under external magnetic field, as shown in Figure 5a. The polarization of fundamental light is rotated by an angle of $\theta_1$, from the solid red arrow to the dashed red arrow. The polarization of generated second harmonic signal is also rotated by an angle of $\theta_2$, which is denoted similarly by the solid and dashed blue arrows. Since the polarizer and analyzer for the magneto-SHG measurements are set at zero magnetic field, we can deduce the formula for the polarization-resolved SHG patterns with extrinsic nonlinear Kerr rotation:

$$E_{2\omega}^{\parallel} \propto \begin{cases} \left[\chi^{(i)}{}_{bbb}\cos^2(\varphi+\theta_1) + \chi^{(i)}{}_{baa}\sin^2(\varphi+\theta_1)\right]\cos(\varphi-\theta_2) \\ +\chi^{(i)}{}_{aab}\sin 2(\varphi+\theta_1)\sin(\varphi-\theta_2) \end{cases} E_{\omega}^2 \quad (9)$$

$$E_{2\omega}^{\perp} \propto \begin{cases} \left[\chi^{(i)}{}_{bbb}\cos^2(\varphi+\theta_1) + \chi^{(i)}{}_{baa}\sin^2(\varphi+\theta_1)\right]\sin(\varphi-\theta_2) \\ -\chi^{(i)}{}_{aab}\sin 2(\varphi+\theta_1)\cos(\varphi-\theta_2) \end{cases} E_{\omega}^2 \quad (10)$$

for WTe$_2$ bilayer, and:

$$E_{2\omega}^{\parallel} \propto \left[\chi^{(i)}{}_{bbb}\cos 3\left(\varphi - \frac{\theta_2 - 2\theta_1}{3}\right)\right] E_{\omega}^2 \quad (11)$$

$$E_{2\omega}^{\perp} \propto \left[\chi^{(i)}{}_{bbb}\sin 3\left(\varphi - \frac{\theta_2 - 2\theta_1}{3}\right)\right] E_{\omega}^2 \quad (12)$$

for WSe$_2$ monolayer, respectively, with $\theta_1 = \mathbf{V_1} \cdot \mathbf{B} = V_1 B$ and $\theta_2 = \mathbf{V_2} \cdot \mathbf{B} = V_2 B$ according to the expression of Faraday rotation. The detailed electric field evolution process is shown in Supporting Information Text 1.

The Verdet constant $V_1$ is obtained by measuring the Faraday rotation angle of the fundamental beam (800 nm) reflected from the bare SiO$_2$/Si surface. As shown in Figure 5b, the polarization of the reflected beam rotates when the magnetic field is swept. The Faraday rotation angle is extracted by fitting the polarization patterns with:

$$I(\theta_1) = I_0 \cos^2(\varphi + 2\theta_1) \quad (13)$$

Note that the rotation angle $\theta_1$ is doubled because of the incident and reflected beams. The extracted Faraday rotation angle as a function of magnetic field is plotted in Figure 5c. By fitting the linear relation with the magnetic field, we obtain $V_1$ of 2.9 °/T.

The Verdet constant $V_2$ is then deduced by fitting the SHG patterns with the fixed zero-field $\chi^{(i)}$ and the obtained $V_1$. As shown in Figure 5d for WTe$_2$ bilayer, the fitted curves well reproduce the data shown in Figure 2e. The obtained magnetic field dependent $\theta_2$ is plotted in Figure 5e. We conduct the similar analysis on WSe$_2$ monolayer and even hBN thin flake next to the WTe$_2$ bilayer. Their magnetic field dependent $\theta_2$ are also plotted in Figure 5e. The three sets of $\theta_2$ from different materials with distinct symmetry are almost overlapped, indicating a common origin from the Faraday rotation effect. Based on the linear relation with magnetic field, the deduced $V_2$ is 11.6 °/T at the frequency-doubled second harmonic wavelength. For optical materials, the Verdet constant is inversely proportional to the square of wavelength.[37] Here the Verdet constant $V_2$ is almost four-times of $V_1$, again suggesting the dominance of Faraday rotation effect. We then conclude that the nonlinear Kerr rotation from these 2D materials is extrinsic and arises from the Faraday rotation effect.

**Extension to 3D topological crystal TaAs**

Beyond 2D systems, we also investigate a typical 3D topological crystal TaAs, which is a Weyl semimetal and highly efficient nonlinear optical material.[23, 25, 28] TaAs is therefore a potential candidate for observing intrinsic nonlinear Kerr rotation in non-magnetic materials. Its atomic structure is shown in Figure 6a, belonging to point group $C_{4v}$. We conducted polarization-resolved SHG on the (112) plane of a TaAs crystal, shown in Figure 6b. The measured polarization-resolved SHG patterns at zero magnetic field are shown in Figure 6c. The two-fold SHG patterns in both XX and XY geometries are consistent with the uniaxially polarized (112) plane of TaAs. Under the electric-dipole approximation, the SHG patterns are fitted to the solid lines in Figure 6c. According to the $C_{4v}$ crystallographic symmetry of TaAs, there are three independent non-zero tensor elements:[23] $\chi^{(i)}_{ccc}, \chi^{(i)}_{caa} = \chi^{(i)}_{cbb}, \chi^{(i)}_{aca} = \chi^{(i)}_{bcb} = \chi^{(i)}_{aac} = \chi^{(i)}_{bbc}$. For its (112) plane, the relation between the input and output electric fields is:

$$E^{\parallel}_{2\omega} \propto \left\{ \begin{matrix} \left[ \chi_{bbb_{eff}} \cos^2(\varphi + \theta_1) + \chi_{baa_{eff}} \sin^2(\varphi + \theta_1) \right] \cos(\varphi - \theta_2) \\ + \chi_{aab_{eff}} \sin 2(\varphi + \theta_1) \sin(\varphi - \theta_2) \end{matrix} \right\} E^2_\omega \quad (14)$$

$$E^{\perp}_{2\omega} \propto \left\{ \begin{matrix} \left[ \chi_{bbb_{eff}} \cos^2(\varphi + \theta_1) + \chi_{baa_{eff}} \sin^2(\varphi + \theta_1) \right] \sin(\varphi - \theta_2) \\ - \chi_{aab_{eff}} \sin 2(\varphi + \theta_1) \cos(\varphi - \theta_2) \end{matrix} \right\} E^2_\omega \quad (15)$$

with $\chi_{bbb\_eff} = \frac{1}{3\sqrt{3}} \left( \chi^{(i)}_{ccc} + 2\chi^{(i)}_{caa} + 4\chi^{(i)}_{aca} \right)$, $\chi_{bbb\_eff} = \frac{1}{\sqrt{3}} \chi^{(i)}_{aca}$ and $\chi_{bbb\_eff} = \frac{1}{\sqrt{3}} \chi^{(i)}_{caa}$.

When the external magnetic field is applied, the XX pattern rotates and gradually decreases (Figure 6d). In the corresponding XY patterns, two of its four lobes shrink and the other two enhance. Although the variation of magneto-SHG patterns is not similar to those in WTe$_2$ bilayer or WSe$_2$ monolayer, we can still reproduce these patterns well with the same $V_1$ and $V_2$ deduced from Figure 5. The simulation results are presented as solid lines in Figure 6d, which match with the magneto-SHG data. Therefore, the observed magneto-SHG is induced by Faraday rotation effect and the intrinsic nonlinear Kerr rotation is still absent.

**Discussion**

We measure and analyze the nonlinear Kerr rotation in several topological crystals by polarization-resolved magneto-SHG. Based on the symmetry related band topology, we propose an intrinsic origin of nonlinear Kerr rotation in topological materials. Yet no such intrinsic effect is observed in the materials we studied. The behavior of nonlinear Kerr rotation rather comes from the Faraday rotation effect through the microscope objective housed inside the superconducting magnet. The extrinsic nonlinear Kerr rotation cannot be simply deducted by a counteracting rotation. For materials with distinct crystallographic symmetry, the nonvanishing second-order nonlinear optical tensor elements are different. Their subtle coupling with Faraday rotation effect leads to different behavior of the extrinsic nonlinear Kerr rotation.

To reduce or avoid the extrinsic nonlinear Kerr rotation, several approaches can be taken. The Faraday rotation effect linearly depends on the applied magnetic field. For an optical setup in Faraday geometry, the magnetic field dependent rotation angles $\theta_1$ and $\theta_2$ can be compensated by properly setting the polarizer and analyzer at each magnetic field. Since the Faraday rotation effect acts on optical components such as transmission-type microscope objective, it can be reduced by replacing with a reflective microscope objective. For a setup in Voigt geometry, where the magnetic field is perpendicular to the wave vector of light, the Faraday rotation effect naturally vanishes and the extrinsic nonlinear Kerr rotation can be circumvented. Moreover, the circularly polarized magneto-SHG measurements are not affected by the Faraday rotation, as shown in Supporting Information Text 2. The extrinsic effect can thus be avoided by measuring magneto-SHG with circularly polarized incident light.

Once the extrinsic nonlinear Kerr rotation diminishes, the band topology induced nonlinear optical effects could be possibly observed. In our experiment, the magnetic field induced energy splitting is relatively small (< 1 meV). The spectral broadening of the femtosecond laser is yet much larger (~10 meV). The intrinsic topological nonlinear optical effect is thus hidden. Perhaps a laser with narrower spectrum is needed. Moreover, if there is some optical resonance with topological non-trivial energy bands, the intrinsic topological effects may be drastically enhanced.

## Author Contributions

S.W.W. conceived and supervised the project. S.W. performed the experiments with assistance from Z.F., Z.S. and Y.Y. Z.F., J.Y., D.H.C and X.X. provided the WTe$_2$ samples. W.X., D.Y., Y.G. and Y.S. provided the TaAs crystals. S.W., Z.F., Z.S., Y.Y., X.X., W.L. and S.W.W. analyzed the data. S.W., Z.S., X.X. and S.W.W. wrote the paper with contributions from all authors.

## Acknowledgments

The work at Fudan University was supported by National Key Research and Development Program of China (Grant Nos. 2019YFA0308404, 2022YFA1403302), National Natural Science Foundation of China (Grant Nos. 12034003, 91950201, 12004077), Science and Technology Commission of Shanghai Municipality (Grant No. 20JC1415900, 2019SHZDZX01, 21JC1402000), Program of Shanghai Academic Research Leader (Grant No. 20XD1400300), Shanghai Municipal Education Commission (2021KJKC-03-61), and China National Postdoctoral Program for Innovative Talents (Grant No. BX20200086). W.X. was supported by the State Key Laboratory of Surface Physics and Department of Physics, Fudan University (KF2022_13). Y.S. thanks the support from the National Natural Science Foundation of China (Grant No. U2032204) and the Informatization Plan of Chinese Academy of Sciences (CAS-WX2021SF-0102).

**Figures and Captions**

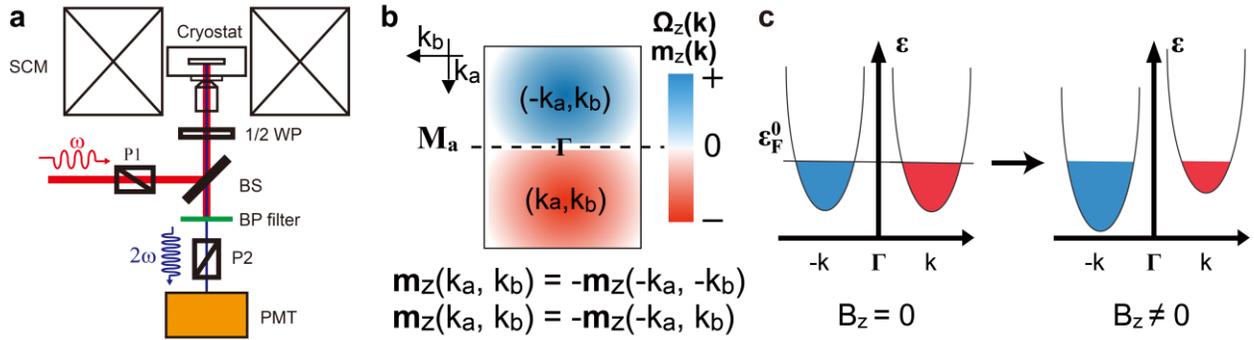

**Figure 1.** Schematics of momentum dependent orbital magnetic moment and experimental setup. (a) Schematic of SHG measurement under external magnetic field. Sample is mounted onto the cryostat placed inside a superconducting magnet (SCM). A microscope objective is used for optical microscopy and SHG imaging. For SHG measurement, the fundamental light with frequency ω is focused to the sample, while the SHG signal with frequency 2ω is collected by a photomultiplier tube (PMT) in photo-counting mode. WP, waveplate; BS, beam splitter; BP, band-pass; P1 and P2, Glan–Thompson polarizers. (b) Orbital magnetic moment ($m_z$) and Berry curvature ($\Omega_z(k)$) are present, but constrained by spatial-inversion and time-reversal symmetries. (c) Illustration of the orbital magnetization under out-of-plane magnetic field.

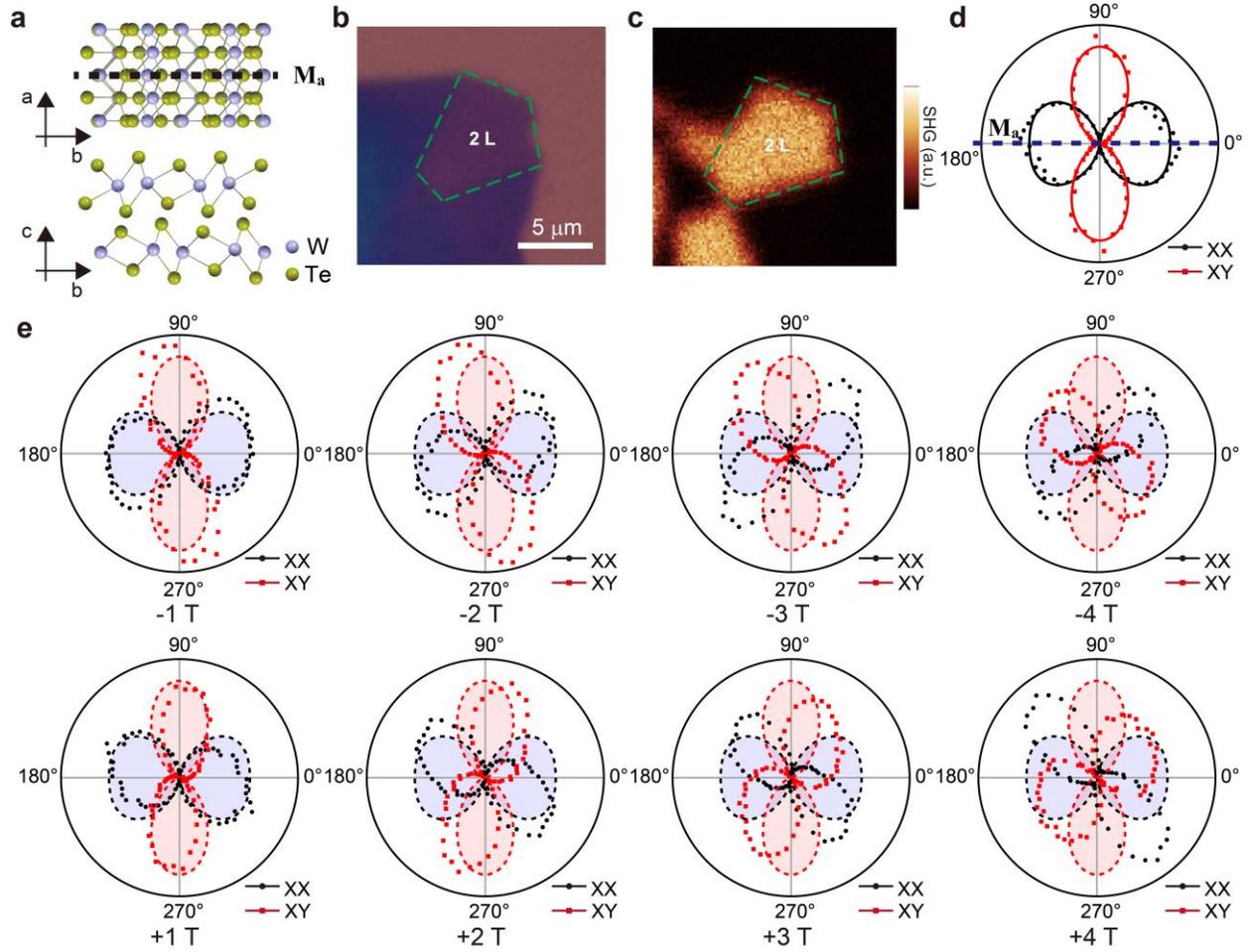

**Figure 2.** SHG response from WTe$_2$ bilayer under out-of-plane magnetic field B. (a) Top (upper panel) and side (lower panel) view of the atomic structures of WTe$_2$ bilayer, which is non-centrosymmetric. The mirror plane of WTe$_2$ bilayer is indicated by the black dashed line. (b) Optical microscopy (before the capping of hBN) of a WTe$_2$ bilayer. (c) The corresponding SHG intensity image of the WTe$_2$ bilayer in (b) at 5 K and B = 0. Here, the fundamental wavelength and excitation power were 800 nm and 0.6 mW, respectively. (d) Polarization-resolved SHG pattern in WTe$_2$ bilayer at B = 0. The excitation and detection are either co- (XX, black dots) or cross- (XY, red dots) linearly polarized. Solid lines are fits to both XX and XY patterns. The zero azimuthal angle lies in the mirror plane **M$_a$** (dashed blue line). (e) Polarization-resolved SHG patterns under negative and positive out-of-plane magnetic field at 5 K. The SHG patterns without magnetic field, as in (d), are shown in blue (XX) and red (XY) shaded area for direct comparisons.

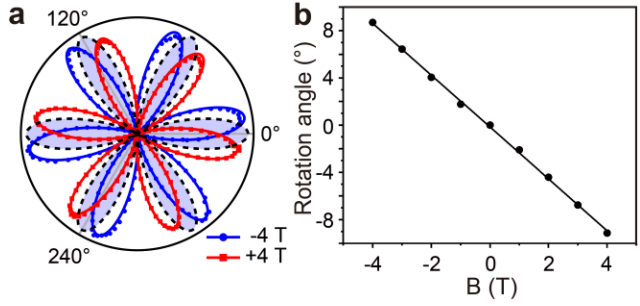

**Figure 3.** SHG response from WSe$_2$ monolayer under out-of-plane magnetic field B. (a) Polarization-resolved SHG patterns of WSe$_2$ monolayer under ±4 T out-of-plane magnetic field at 5 K. The SHG patterns without magnetic field are shown in blue shaded area for direct comparison. For clarity, only the XX patterns are shown. (b) Rotation angle extracted from the SHG patterns under different out-of-plane magnetic field. The black points are the extracted angle values and the straight line is the linear fit.

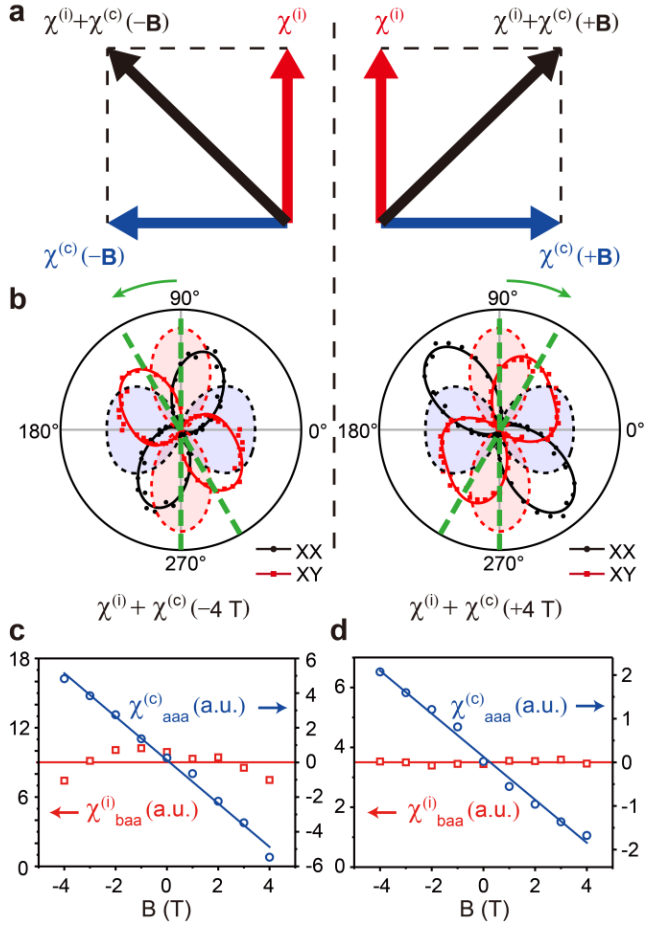

**Figure 4.** Nonlinear Kerr rotation induced by superposition of $\chi^{(i)}$ and $\chi^{(c)}$. (a) Schematic of nonlinear Kerr rotation under negative (left panel) and positive (right panel) magnetic field. (b) SHG polarization patterns rotating to the opposite directions under -4 T (left panel) and +4 T (right panel) out-of-plane magnetic field, respectively. The solid curves are the fitting. The shaded areas are the corresponding zero-magnetic-field SHG patterns. (c, d) i-type (red open squares) and c-type (blue open circles) second-order nonlinear tensor elements for WTe$_2$ bilayer (c) and WSe$_2$ monolayer (d), obtained by fitting the SHG patterns under different magnetic field. Lines are guides to the eye.

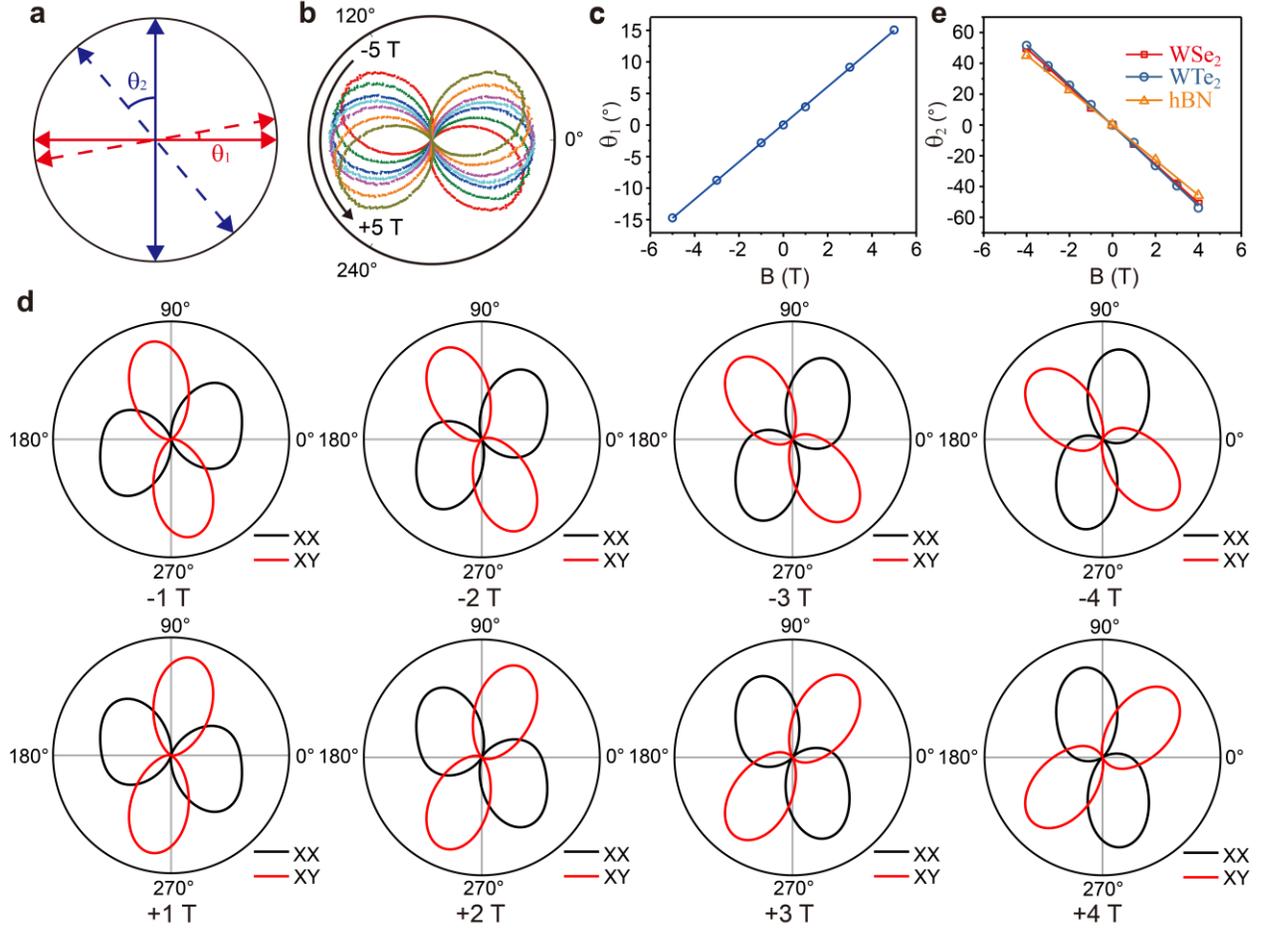

**Figure 5.** Faraday rotation effect induced extrinsic nonlinear Kerr rotation. (a) Schematic of Faraday rotation angle of the fundamental ($\theta_1$) and second harmonic ($\theta_2$) light. (b) Polarization patterns of reflected light (wavelength, 800 nm) measured on the substrate under -5 T to 5 T out-of-plane magnetic field. (c) Faraday rotation angle of the fundamental light fitted and extracted from (b). (d) Fitting results of polarization-resolved SHG patterns of WTe$_2$ bilayer under different out-of-plane magnetic field. (e) Faraday rotation angle $\theta_2$ of different materials extracted from the magneto-SHG polarization patterns.

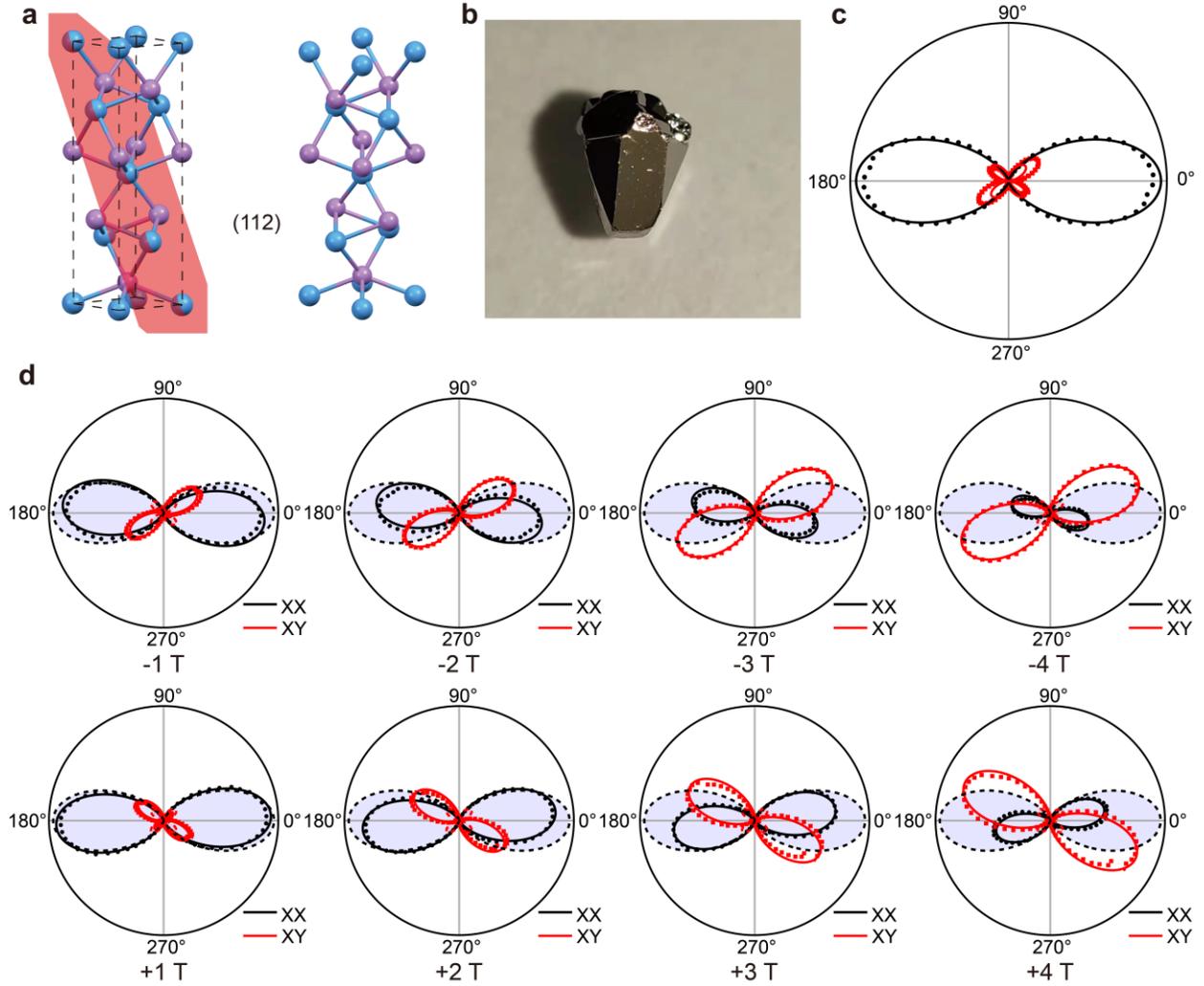

**Figure 6.** Extrinsic nonlinear Kerr rotation in TaAs (112). (a) Atomic structure of TaAs crystal (left panel) and viewed from (112) plane (right panel), respectively. The (112) plane is the red plane in left panel. (b) TaAs crystal used in SHG measurement. (c) Polarization-resolved SHG pattern in TaAs (112). Solid lines are fits to both XX and XY patterns. (d) Polarization-resolved SHG patterns under negative and positive out-of-plane magnetic field at 5 K. The SHG patterns without magnetic field, as in (c), are shown in blue (XX) and red (XY) shaded area for direct comparisons. Solid lines are simulations of the magneto-SHG patterns.

# Supporting Information

**Supplementary Text**

1. **Jones matrices of the optical components and electric field evolution process of the extrinsic nonlinear Kerr rotation**

   The Jones matrices of the key optical component in our experiment, the microscope objective, is $A_1 = \begin{pmatrix} \cos\theta_1 & \sin\theta_1 \\ -\sin\theta_1 & \cos\theta_1 \end{pmatrix}$ for the incident beam and $A_2 = \begin{pmatrix} \cos\theta_2 & \sin\theta_2 \\ -\sin\theta_2 & \cos\theta_2 \end{pmatrix}$ for the second harmonic beam, with $\theta_1 = V_1 B$, $\theta_2 = V_2 B$, $V_1 = 2.9$ °/T and $V_2 = 11.6$ °/T.

   The incident beam can be presented as $\mathbf{E_0} = \begin{pmatrix} E_\omega \cos(\varphi + \theta_0) \\ E_\omega \sin(\varphi + \theta_0) \end{pmatrix}$. After going through the microscope objective, the electric field of the beam evolves to

   $$\mathbf{E'_0} = A_1 \mathbf{E_0} = \begin{pmatrix} E_\omega \cos(\varphi + \theta_0) \cos\theta_1 + E_\omega \sin(\varphi + \theta_0) \sin\theta_1 \\ -E_\omega \cos(\varphi + \theta_0) \sin\theta_1 + E_\omega \sin(\varphi + \theta_0) \cos\theta_1 \end{pmatrix}$$

   whose polarization direction is rotated by an angle of $\theta_1$.

   The electric field of the generated second harmonic beam is

   $$\mathbf{E_{2\omega}} = \vec{\chi} : \mathbf{E'_0} \mathbf{E'_0} = \begin{pmatrix} E_{2\omega} \cos(\varphi + \theta_0) \\ E_{2\omega} \sin(\varphi + \theta_0) \end{pmatrix}$$

   where $\vec{\chi}$ depends on the crystallographic symmetry of the specific sample. After going through the microscope objective, the electric field can be written as

   $$\mathbf{E'_{2\omega}} = A_2 \mathbf{E_{2\omega}} = \begin{pmatrix} E_{2\omega} \cos(\varphi + \theta_0) \cos\theta_2 + E_{2\omega} \sin(\varphi + \theta_0) \sin\theta_2 \\ -E_{2\omega} \cos(\varphi + \theta_0) \sin\theta_2 + E_{2\omega} \sin(\varphi + \theta_0) \cos\theta_2 \end{pmatrix}$$

   whose polarization direction is rotated by an angle of $\theta_2$.

2. **Electric field evolution process with a circularly polarized incident beam**

   With a circularly polarized incident beam, the Jones matrices is the same as the linearly polarized incidence. The electric field, however, should be presented as: $\mathbf{E_0}^L = \frac{1}{\sqrt{2}} \begin{pmatrix} E_\omega \\ E_\omega e^{\frac{\pi}{2} i} \end{pmatrix}$ for left-handed circularly polarized incident beam and $\mathbf{E_0}^R = \frac{1}{\sqrt{2}} \begin{pmatrix} E_\omega \\ E_\omega e^{-\frac{\pi}{2} i} \end{pmatrix}$ for left-handed circularly polarized incident beam, respectively. After going through the microscope objective, the electric field of the beam evolves to

$$\mathbf{E'_0}^L = \mathbf{A_1} \mathbf{E_0}^L = \frac{1}{\sqrt{2}} \begin{pmatrix} E_\omega \cos\theta_1 + E_\omega e^{\frac{\pi}{2}i} \sin\theta_1 \\ -E_\omega \sin\theta_1 + E_\omega e^{\frac{\pi}{2}i} \cos\theta_1 \end{pmatrix}$$

$$= \frac{1}{\sqrt{2}} \begin{pmatrix} E_\omega \\ E_\omega e^{\frac{\pi}{2}i} \end{pmatrix} \left( \cos\theta_1 + e^{\frac{\pi}{2}i} \sin\theta_1 \right)$$

$$= \frac{1}{\sqrt{2}} \begin{pmatrix} E_\omega \\ E_\omega e^{\frac{\pi}{2}i} \end{pmatrix} e^{\theta_1 i}$$

and

$$\mathbf{E'_0}^R = \mathbf{A_1} \mathbf{E_0}^R = \frac{1}{\sqrt{2}} \begin{pmatrix} E_\omega \cos\theta_1 + E_\omega e^{-\frac{\pi}{2}i} \sin\theta_1 \\ -E_\omega \sin\theta_1 + E_\omega e^{-\frac{\pi}{2}i} \cos\theta_1 \end{pmatrix}$$

$$= \frac{1}{\sqrt{2}} \begin{pmatrix} E_\omega \\ E_\omega e^{-\frac{\pi}{2}i} \end{pmatrix} \left( \cos\theta_1 + e^{-\frac{\pi}{2}i} \sin\theta_1 \right)$$

$$= \frac{1}{\sqrt{2}} \begin{pmatrix} E_\omega \\ E_\omega e^{-\frac{\pi}{2}i} \end{pmatrix} e^{-\theta_1 i}$$

for left- and right-handed circularly polarized incident beams, respectively. The resulting light is still left- and right-handed circularly polarized but with their phases shifted by values of $\theta_1$ and $-\theta_1$, respectively. The phase shift of circularly polarized light will not affect the SHG intensity. Therefore, the SHG result with left- and right-handed circularly polarized incident light in a magnetic field has no difference.